
\input harvmac.tex

\def\LG{Lan\-dau-Ginz\-burg\ }
\def\gkl{G_{k,\ell}}
\def\mkl{{\cal M}_{k,\ell} (G;H)}
\def\Gkl#1{{{G_k \times #1_\ell} \over {#1_{k+\ell}}}}
\def\UU{{\cal U}} \def\half{{1 \over 2}}
\def\primg{\Phi^{\Lambda, \Lambda_+}_{\Lambda_-}}
\def\primh{\Phi^{\Lambda, \lambda_+}_{\lambda_-}}
\def\nup#1{{\it Nucl.\ Phys.} \ {\bf B#1\/}}
\def\ijmp#1{{\it Int.\ J. \ Mod. \ Phys.} \ {\bf A#1\/}}
\def\plt#1{{\it Phys.\ Lett.}\ {\bf#1B\/}}
\def\cmp#1{{\it Commun. \ Math. \ Phys.} \ {\bf #1\/}}
\def\scr{{\cal R}} \def\scf{{\cal F}} \def\Gminus{G_{-{1 \over 2}}^-}
\def\Gplus{G_{-{1 \over 2}}^+}
\def\coeff#1#2{\relax{\textstyle {#1
\over #2}}\displaystyle}
\def\inbar{\vrule height1.5ex width.4pt depth0pt}
\def\IC{\relax\,\hbox{$\inbar\kern-.3em{\rm C}$}}
\def\IP{\relax{\rm I\kern-.18em P}}
\def\IR{\relax{\rm I\kern-.18em R}}
\font\sanse=cmss12
\def\ZZ{\relax{\hbox{\sanse Z\kern-.42em Z}}}

%
%
\def\Titletwo#1#2#3#4{\nopagenumbers\abstractfont\hsize=\hstitle
\rightline{#1}\rightline{#2} \vskip .7in\centerline{\titlefont #3}
\vskip .1in \centerline{\abstractfont {\titlefont #4}}
\abstractfont\vskip .5in\pageno=0}

\Titletwo{}{} {Topological Matter, Integrable Models} {and Fusion
Rings$^*$ {\abstractfont \footnote{}{$^*$ Work supported in part by
funds provided by the DOE under grant No. DE-FG03-84ER40168.}} }
\centerline{D. Nemeschansky \ and \ N.P. Warner \footnote{$^\dagger$}{
Alfred P. Sloan foundation fellow.}} \bigskip \centerline{Physics
Department} \centerline{University of Southern California}
\centerline{University Park} \centerline{Los Angeles, CA 90089-0484.}
\vskip 1.0cm
We show how topological $G_k/G_k$ models can be embedded
into the topological matter models that are obtained by perturbing the
twisted $N=2$ supersymmetric, hermitian symmetric, coset models.  In
particular, this leads to an embedding of the fusion ring of $G$ as a
sub-ring of the perturbed, chiral primary ring.  The perturbation of
the twisted $N=2$ model that leads to the fusion ring is also shown to
lead to an integrable $N=2$ supersymmetric field theory when the
untwisted $N=2$ superconformal field theory is perturbed by the same
operator and its hermitian conjugate.
\vfill
\leftline{USC-91/031 }
\Date{October, 1991}

\newsec{Introduction}

The structure of the fusion ring, or Verlinde algebra, for affine Lie
algebras has been known for quite some time \ref\KW{V.G.~Kac, talk at
the Canadian Mathematical Society Meeting on Lie algebras and Lie
groups, CRM, Universite
 de Montreal, Aug. 89, unpublished; M.~Walton, \nup{340} (1990) 777; M.
{}~Walton, \plt{241} (1990) 365;  J.~Fuchs and P.~van~Driel, ``WZW fusion
rules, quantum groups  and the modular invariant matrix S'', preprint
ITFA-90-03 (1990); P.~Furlan, A.~C~Ganchev and V.B.~Petkova, \nup{343}
(1990) 205; M.~Spiegelglas, \plt{247} (1990) 36;  P.\ Bouwknegt,
J.~McCarthy, D.~Nemeschansky and K.~Pilch, \plt{258} (1991) 127.}
 \ref\EV{ E.~Verlinde, \nup{300} (1988) 360.}.  Perhaps one of the
simplest characterizations of this fusion ring comes from the
``discrete  characters'' \EV, whose multiplication table precisely
reproduces the fusion rules.  More recently, it was shown in
\ref\Gep{D.~Gepner, ``Fusion Rings and Geometry'', Santa Barbara
preprint, NFS-ITP-90-194 (1990); ``On the Algebraic Structure of $N=2$
String Theory'', Weizmann preprint WIS-90-47-PH (1990).}
 that the fusion ring of $SU_k(n)$ could be ``integrated'' to yield a
potential $W(x_a; \lambda)$, whose local ring is isomorphic to the
fusion ring.  That is, the fusion ring could be characterized by a
single function $W(x_a; \lambda)$: the variables, $x_a$, are
canonically identified with particular representations of $SU_k(n)$,
and the algebraic relations implied by the fusion rules consist
precisely of those polynomials of $x_a$ that lie in the ideal generated
by the partials $\big\{ {{\partial W} \over {\partial x_a}} \big\}$.
At the time, it was unclear whether this remarkable result could be
generalized.

The potentials $W(x_a; \lambda)$ also turned up in an apparently
unrelated piece of work on the structure of a generalized Zamolodchikov
metric \ref\AZM{ A.~Zamolodchikov, {\it JETP Lett.} {\bf 43} (1986)
731.} for the ground states  of the perturbed $N=2$ coset models that
have a \LG structure.  It was shown in \ref\CeVa{S.~Cecotti and
C.~Vafa, ``Topological Anti-Topological Fusion'', Harvard preprint
HUTP-91/A031 (1991).} that for a special class of perturbations, this
generalized Zamolodchikov metric satisfied some classical Toda
equations.  This indirectly suggested that the corresponding perturbed
conformal models should be integrable.  For the $\IC \IP^n$ models,
this particular class of perturbations gave rise to an effective \LG
potential, $W(x_a; \lambda)$, that was identical to the potential that
characterized the fusion rules of $SU_k(n)$.

Further suggestive results have been obtained in
\ref\MSG{M.~Spiegelglas,
 and S.~Yankielowicz,`` $G/G$ Topological Field Theory  by Coseting'',
in preparation; ``Fusion rules as Amplitudes in $G/G$ Theories'', in
prepartion.} and
 \ref\MSS{M.~Spiegelglas, ``Setting Fusion rings in topological Landau
Ginzburg  theories'', preprint Technion-PH-8-91 (1991).}.
Unfortunately, the first of these two papers had not appeared at the
time of writing this paper, but one of the central ideas of
\MSG\ appears to be that the Verlinde algebra of $G_k$ can be obtained
from the correlation functions of a topological coset model based on
$G_k/G_k$.  The idea in \MSS\ is that such a fusion algebra might
possibly be naturally identified with a particular topological
perturbation of the chiral primary ring \ref\DVV{R.~Dijkgraaf,
E.~Verlinde and H.~Verlinde, \nup{352} (1991) 59.} of a suitably chosen
twisted $N=2$ superconformal field theory.  The $SU(2)_k$ example is
extensively discussed in \MSS, but a complete understanding of the type
$D$ and type $E$ modular invariants appeared elusive.  It is however
fairly clear from the results in \Gep\ that something like this should
be true for $SU_k(n)$.

The conjecture in \MSG\ about $G/G$ models was recently established in
\ref\witt{E.~Witten, ``Holomorphic factorization of WZW coset models'',
Princeton preprint, IASSNS-HEP-91-25 (1991).}.  Direct evidence for the
connection with topological perturbations of $N=2$ \LG models can be
found in \ref\KI{K.~Intriligator, ``Fusion residues'', Harvard preprint
HUTP-91/A041 (1991).}, where the Verlinde dimensions for $G_k$ theories
are expressed as correlation functions in topological \LG theories.

Our purpose in this paper is to construct directly the embeddings of
$G_k/G_k$ theories into the topological matter theories that are
obtained by twisting hermitian symmetric space, $N=2$ supersymmetric,
coset models.  Specifically, the correlation functions of $G_k/G_k$ can
all be translated directly into a subset of the topological correlation
functions of a twisted $N=2$ supercoset theory.  In particular, the
fusion algebra of $G_k$ generally forms a natural sub-ring of the
perturbed chiral, primary ring, $\scr$.  We describe this embedding in
detail in section 3 of this paper.  To obtain the $G_k/G_k$ theory from
the $N=2$ theory one perturbs away from the conformal point in a very
specific, flat, or geodesic, direction.  We identify this direction and
the corresponding relevant perturbing operator.  Naturally, for
$SU_k(n)$ we recover the results of \Gep\ and \KI.

In section 4, we show that if the original $N=2$ superconformal field
theory (prior to twisting) is perturbed by the operator considered
above, along with the hermitian conjugate perturbation, then the result
is an integrable field theory.  In particular, we establish that a
general Grassmannian model has at least three integrable
perturbations.

The key to establishing all of these results is essentially contained
in the original paper by Eguchi and Yang \ref\EY{T.~Eguchi and
S.-K.~Yang, {\it Mod. \ Phys. \ Lett.} {\bf A5} (1990) 1693. }, where
the group $SU_k(2)$ is considered.  The generalization of \EY\ is
relatively straightforward once one has obtained the appropriate
description of the $N=2$ supersymmetric, hermitian symmetric space
models.  For reasons that will soon become obvious, we call this the
{\it paratoda } description, and we will discuss it in section 2.

The paratoda construction is not manifestly supersymmetric.  However,
for the $\IC \IP ^n$ supercoset models, there is a free superfield
realization that can be obtained by hamiltonian reduction
\ref\HN{D.~Nemeschansky and S.~Yankielowicz,``$N=2$ W-algebras,
Kazama-Suzuki Models and Drinfeld-Sokolov Reduction'', USC-preprint
USC-007-91 (1991)}, \ref\KIto{K.~Ito, \plt{259} (1991) 73.}. In this
formulation, the perturbed model is easily seen to be integrable and
the top components of the supermultiplets can be seen to provide the
quantum integrals of motion.  This is discussed in section 5.

\newsec{The Parafermionic-Toda Models}

Our purpose in this section is to show how to represent an $N=2$
supersymmetric coset model based upon a hermitian symmetric space (HSS)
in terms of an appropriately screened tensor product of generalized
parafermions and a Toda theory (or, at least, the free field equivalent
of a Toda theory).  We will, in fact, obtain such a {\it paratoda}
description for the slightly more general class of coset models:
$$\mkl ~\equiv~ \Gkl{H} \ ,$$
where $H$ is a subgroup of $G$, with
rank$(H)$ $=$ rank$(G)$ \foot{ It probably suffices to merely have $H$
be a regularly embedded subgroup of $G$.}.
Let $\alpha_1, \ldots , \alpha_r$ be a system of simple roots for $G$,
ordered in such a way that $\alpha_1, \ldots , \alpha_p$ is a system
of simple roots for $H$.  Let $\UU = (U(1))^r$ be a torus for $H$, and
hence a torus for $G$.

First consider the situation where $H \equiv G$.  Such models will
be called $\gkl$ coset models.  These have central charge  $c =
c_k(G) + c_\ell(G) -c_{k + \ell}(G)$, where $c_m(G)$ is the central
charge of $G$ at level $m$.  The central charge of the $\gkl$ model may
be written \ref\BL{C.~Ahn, D.~Bernard and A. \ LeClaire, \nup{346}
(1990) 409.}:
\eqn\centch{c ~=~ \bigg[ {{k d_G} \over {(k+g)}} ~-~ r
{}~\bigg] ~+~ \bigg[~ r {}~-~ {{12 k \rho_G^2} \over {(\ell+g)(k + \ell
+g)}} \bigg] \ ,}
where $d_G$ is the dimension of $G$, $g$ is the
dual Coxeter number of $G$, and $\rho_G$ is the Weyl vector of $G$.
The first term in \centch\ is the central charge of the generalized
parafermionic coset $G/\UU$, and the second term is the central charge
of a  Feigen-Fuchs, free bosonic field theory, or of its equivalent
Toda theory.  It has been convincingly argued in \ref\BN{J.~Bagger and
D.~Nemeschansky and S.~Yankielowicz, Phys. Rev. Lett. {\bf 60} (1988)
389. }, \ref\KMQ{D.~Kastor, E.~Martinec and Z.~Qiu, \plt {226} (1988)
434.} \ref\Nam{S.~Nam, \plt{243} (1990)  231.}, \BL\ that the $\gkl$
coset theory can indeed be obtained from  a tensor product of such
theories.  Specifically, one tensors the parafermionic theory $G/\UU$
with a theory consisting of $r$ free bosons with energy-momentum
tensor:
$$T_{b}(z) ~=~ - \half (\partial \phi )^2 ~+~ i ~(\beta_+
{}~-~ \beta_-) \rho_G \cdot \partial ^2 \phi \ ,$$
where
\eqn\betadef{\beta_\pm ~\equiv~ {1 \over \sqrt{k}} ~ \Bigg[
\sqrt{{{(k+\ell + g)} \over {(\ell + g)}}}~ \Bigg]^{\pm 1}\ .}
As has
been described in a number of places, such a bosonic free field
description can be directly related to a Toda theory
(see, for example,
\ref\BilGe{A.~Bilal and  J.-L. \ Gervais, \plt{206} (1988) 412;
\nup{318} (1989) 579; \nup{326} (1989) 222;
A.\ Bilal, \nup{330} (1990)
399;  \ijmp{5} (1990) 1881.}
\ref\HM{T.~Hollowood and P.~Mansfield,
\plt {226} (1989) 73.}).

The primary fields of the parafermionic theory will be denoted by
${\cal A}^\Lambda_\lambda$, where $\Lambda$ is a highest weight of
$G_k$, and $\lambda$ is a vector of charges under the Cartan subalgebra
(CSA), ${\cal X}$, of $G$ that generates the torus, $\UU$.  The Toda,
or free bosonic,  field theory has a natural vertex operator
representation for its highest weight states:
\eqn\todavert{{\cal
V}_{\Lambda_+, \Lambda_-}(z) ~=~ exp \big[ - i (\beta_+ \Lambda_+ ~-~
\beta_- \Lambda_- ) \cdot \phi \big ] \ .}
The conformal weight  of
${\cal A}^\Lambda_\lambda$ is:
$$ h^\Lambda_\lambda ~=~ {{\Lambda
\cdot (\Lambda + 2 \rho_G ) } \over {2 (k + g)} }  ~-~ {{\lambda^2}
\over {2k}} ~+~ {\rm integer} \ ,$$
 and that of ${\cal V}_{\Lambda_+,
\Lambda_-}$ is:
$$ h_{\Lambda_+, \Lambda_- }  ~=~ \half \big(
\beta_+ \Lambda_+ ~-~ \beta_- \Lambda_- \big )^2 ~+~ (\beta_+ ~-~
\beta_-) \rho_G \cdot (\beta_+ \Lambda_+ {}~-~ \beta_- \Lambda_- ) \ .
$$
One finds that this can be rewritten as:
\eqn\confwt{h_{\Lambda_+, \Lambda_- }  ~=~  {1 \over {2k}} ~(\Lambda_+
- \Lambda_- )^2 ~+~ {{\Lambda_+ \cdot (\Lambda_+ + 2 \rho_G ) } \over
{2 (\ell + g)} } ~-~ {{\Lambda_- \cdot (\Lambda_- + 2 \rho_G ) } \over
{2 (k + \ell + g)} } \ .}
{}From this, and a consideration of the
Cartan subalgebra eigenvalues, $\lambda$, it is easy to identify
representatives of the primary fields, $\primg$, of the $\gkl$ coset
model. ( The labels $(\Lambda, \Lambda_-; \Lambda_- )$ are highest
weight labels of affine $G$ at levels $k$, $\ell$ and $k+\ell$
respectively, and correspond to the numerator and denominator factors
in $\gkl$.)  Indeed, we may take
\eqn\primfield{\primg(z)~=~{\cal
A}^\Lambda_{(\Lambda_- - \Lambda_+)}(z) {}~{\cal
V}_{\Lambda_+,\Lambda_-}(z)\ .}
 Finally, define the operators:
\eqn\screen{\eqalign{S_{+\alpha_i}(z) ~=~& \Phi^{0, -\alpha_i}_0 (z)
{}~\equiv~ {\cal A}^0_{\alpha_i}(z) ~ exp\big[ + i \beta_+ ~\alpha_i
\cdot \phi \big] \cr S_{-\alpha_i}(z) ~=~& \Phi^{0, 0}_{- \alpha_i} (z)
{}~\equiv~ {\cal A}^0_{- \alpha_i}(z) ~ exp\big[ - i \beta_- ~\alpha_i
\cdot \phi \big] \ ,}}
for $i = 1, \ldots, r$.  These operators have
conformal weight equal to one, and constitute the screening operators
for the theory \BN\KMQ\Nam .

While the correspondence of the $G_{k,\ell}$ coset models with the
foregoing paratoda description has not been rigorously established,
there is a fairly compelling body of evidence.  First, the model with
$G=SU(2)$ has been extensively discussed in \BN\Nam, where the
correspondence was further verified by checking characters.  The
correspondence for general $G$  has been discussed in \KMQ\BL\Nam\ and
it is also fairly clear that the arguments in \BN\ can be extended
beyond $SU(2)$.  Secondly, for the $k=1$ models, $G_{1,\ell}$, the free
field realization given above reduces to the well known one
\ref\FatL{V.~Fateev and V.~Lukyanov, Int. J. Mod. Phys. {\bf A3} (1988)
507. }\HM \ref\BMP{P.~Bouwknegt, J.~McCarthy and K.~Pilch,\nup{352}
(1991) 139.}.  The rigorous proof that this free bosonic realization
yields the $G_{1,\ell}$ models may be found in \BMP.  It is also
probable that this proof could be extended to cover the paratoda
theories \ref\KP{K. \ Pilch, private communication.} .  We will
therefore assume that the paratoda theories with the screening currents
\screen\ describe the $G_{k, \ell}$ models.

It is elementary to generalize this description to the $\mkl$ models.
Observe that
\eqn\Hptmodel{\eqalign{\mkl ~=~ & {G \over {H_k}}
{}~\times~ {{H_k \times H_\ell} \over {H_{k + \ell}}} \cr ~=~ & {G \over
{H_k}} ~\times~ {{H_k } \over {\cal U}} ~\times~ {\cal U} ~=~ {G \over
{\cal U}} ~\times~ {\cal U} \ ,}}
where the second equality comes
from replacing the $H_{k, \ell}$ theory by its paratoda equivalent.
Thus the $\mkl$ models can be obtained by tensoring the parafermionic,
${G \over {\cal U}}$, theory with an $H$-Toda theory.  The
energy-momentum tensor of the $H$-Toda theory is
\eqn\Htoda{T_{b}^\prime (z) ~=~ -\half ~ \big( \partial \phi \big)^2
{}~+~ i~ \big( \beta_+^\prime ~-~ \beta_-^\prime \big) ~ \rho_H \cdot
\partial^2 \phi \ ,}
where $\rho_H$ is the Weyl vector of $H$, and
the coupling constants are given by
\eqn\betaprm{\beta_\pm^\prime
{}~\equiv~ {1 \over \sqrt{k}} ~ \Bigg[ \sqrt{{{(k+\ell + h)} \over {(\ell
+ h)}}}~ \Bigg]^{\pm 1}\ ,}
where $h$ is the dual Coxeter number of
$H$ \foot{ As usual, if $H$ has several factors, the levels, $\ell$ or
$k+\ell$, and the dual Coxeter number, $h$, are to be interpretted as
vectors with entries corresponding to each such factor.  The dual
Coxeter number of a $U(1)$ factor is defined to be zero.}.  The
screening currents are still given by \screen, but with $i$ restricted
to the range $i= 1, \ldots, p\,$; that is, the  $\alpha_i$, are
restricted to the simple roots of $H$.  Representatives of the highest
weight states of the model are still given by \primfield, but with
$\beta_\pm$ replaced by $\beta_\pm^\prime$.  We will denote these
representatives by $\primh (z)$, where $\lambda_+$ and $\lambda_-$ are
now highest weight labels of affine $H$ at level $\ell$ and level $k +
\ell$ respectively. The conformal weight of $\primh$ is given by
\eqn\phiwt{h ~=~ {{\Lambda \cdot (\Lambda + 2 \rho_H ) } \over {2 (k +
g)} } {}~+~ {{\lambda_+ \cdot (\lambda_+ + 2 \rho_H ) } \over {2(\ell +
h)} } ~-~ {{\lambda_- \cdot (\lambda_- + 2 \rho_H ) } \over {2 (k +
\ell + h)} } \ .}

If $H$ has any $U(1)$ factors, then they give rise directly to
uncorrupted free $U(1)$ factors in the bosonic torus theory, ${\cal
U}$, in \Hptmodel.  Such $U(1)$ factors in ${\cal U}$ are uncorrupted
in the sense that they are orthogonal to all screening currents and to
the change at infinity.  Indeed, if $p(z)$ denotes the vector of
currents in the CSA, ${\cal X}$, of $H$, and $v \cdot p(z)$ defines
some $U(1)$ factor in $H$, then the corresponding $U(1)$ factor in
${\cal U}$ is given by $v \cdot \partial \phi$, up to some
normalization.

If $G/H$ is a symmetric space, then $H_{g-h}$ can be conformally
embedded in $SO({\rm dim}({G\over H}))$, and so for a special choice of
modular invariant for $H_{g-h}$,
the $M_{k,\ell=g-h}(G;H)$ model is precisely the
super-GKO coset model based on $G/H$.   If $G/H$ is a hermitian
symmetric space (HSS), then $M_{k,\ell=g-h}(G;H)$ is, of course, an
$N=2$ supersymmetric model
 \ref\KS{Y.~Kazama and H.~Suzuki, \nup{234} (1989) 232. }.  We now
specialize to these $N=2$, HSS models.

By rewriting the construction of \KS\ in terms of the paratoda theory,
one finds that the $U(1)$ current of the
$N=2$ superalgebra is given by:
\eqn\superuone{J(z)
{}~=~ 2i~ ( \beta_+^\prime ~-~ \beta_-^\prime ) ~ (\rho_G {}~-~ \rho_H)
\cdot \partial \phi \ , }
and the associated charge, $Q$, of the
primary field $\primh$ is:
\eqn\Qcharge{ Q ~=~ -2 (\rho_G ~-~
\rho_H) \cdot \bigg[ {{\lambda_+} \over g} {}~-~ {{\lambda_-} \over {k
+ g}} \bigg ] \ .}

In \ref\LVW{W.\ Lerche, C.\ Vafa and N.P.\ Warner, \nup{324} (1989)
427} the chiral, primary fields were identified by computing the Ramond
ground states.  It was shown that there were
\eqn\crdim{\mu ~=~ {1
\over {\vert Z(G) \vert}} ~ N_k(G) ~\bigg \vert {{W(G)} \over {W(H)}}
\bigg \vert }
such states, where $N_k(G)$ is the number of affine,
highest weight labels of $G$ at level $k$, $Z(G)$ is the center of $G$,
and $W(G)$ and $W(H)$ are the (finite) Weyl groups of $G$ and $H$,
respectively.  For each highest weight label, $\Lambda$, of $G$, and
for each highest weight label, $\lambda$, of the coset denominator,
$H_{k + g - h}$, one obtains a Ramond ground state for the coset if and
only if:
\eqn\LLreln{ \lambda ~=~ w(\Lambda ~+~ \rho_G) ~-~ \rho_H
\ , }
where $w$ is a representative of a Weyl-coset element of
$W(G)/W(H)$.  Moreover, one obtains exactly one such ground state for
each such $\Lambda$ and $\lambda$.  The Weyl element $w$ in \LLreln\ is
chosen so that $w(\Lambda + \rho_G)$ is a dominant weight of $H$.  This
choice also guarantees that $\lambda$ is a dominant weight of $H$.  The
weight $w(\Lambda)$ is manifestly a weight in the $G$ representation
with highest weight $\Lambda$, and it is straightforward to establish
that  $w(\Lambda)$ is also a highest weight of affine $H$ at level
$k$.  Similarly $w(\rho_G)-\rho_H$ is a highest weight label for
$H_{g-h}$ and can be obtained from the spinor ground-state of $SO({\rm
dim}(G/H))$.  Thus we can construct representatives of the Ramond vacua
directly from the ground states of the $G$ and $H$ factors in the coset
model.  To obtain representatives of the chiral primary fields from
this description of the Ramond ground states, one merely performs a
spectral flow with respect to the $N=2$ $U(1)$ current, $J(z)$.  If one
simultaneously performs a spectral flow in the CSA of the denominator,
$H$, one can arrange that the net effect of both these spectral flows
is a single spectral flow in the $SO({\rm dim}(G/H))$ factor.  This
means that one can obtain the chiral primary fields by taking the
affine $G$ and $H$ labels  corresponding to the Ramond ground states,
and merely shifting the $H$-labels by $-(\rho_G - \rho_H)$ \foot{The
negative sign comes from our reversal of the conventional sign for
$J(z)$ used in \KS.}.  As a consequence, a set of representatives of
the chiral primary fields is given by:
\eqn\cplabels{\Phi^{\Lambda,
\lambda_+}_{\lambda_-} \ , \qquad {\rm where} \qquad \lambda_+ ~=~
w(\rho_G) - \rho_G\ ; \ \ \lambda_- {}~=~ w(\Lambda + \rho_G) - \rho_G
\ .}

One can also establish this result directly from our earlier analysis
by using \Qcharge\ and \phiwt\ (with $\ell = g - h$) to write:
\eqn\hminusQ{h ~-~ \half Q ~=~ {{\big( \vert \Lambda + \rho_G \vert^2
{}~-~ \vert \lambda_- + \rho_G \vert^2 \big)} \over {2(k+g)}} ~+~ {{\big(
\vert \lambda_+ + \rho_G \vert^2 ~-~ \vert  \rho_G \vert^2 \big)} \over
{2g}} \ .}
It is obvious that the labels in \cplabels\ satisfy $h -
\half Q = 0$, and the reverse implication may be established by the
arguments in the appendix of \LVW, whose crucial observation is that
weights in a $G$-representation have the same length as the highest
weight if and only if they are Weyl images of that highest weight.

The field identifications induced by spectral flow in the CSA, ${\cal
X}$, of $H$, map a coset state with weights $(\Lambda, \lambda_+ ;
\lambda_-)$ into another such state according to:
\eqn\flow
{\eqalign{\Lambda ~\rightarrow~ & \Lambda ~+~ k ~ v \cr \lambda_+
{}~\rightarrow~ & \lambda_+ ~+~ (g - h)  v \cr  \lambda_{-}
{}~\rightarrow~ &  \Lambda_{-} ~+~  (k+g-h) v  }}
where $v$ is any
vector.   Spectral flow by an arbitrary vector, $v$, yields an
automorphism of the coset theory provided we use appropriately twisted
Kac-Moody currents (see, for example, \ref\GodOl{P. \ Goddard and D.
\ Olive, \ijmp{1} (1986) 303.}); that is, we replace the currents of
$G$ or $H$ according to:
$$\eqalign{J^{\, \alpha}_n ~\rightarrow~&
J^{\, \alpha}_{n + v\cdot \alpha} \cr H^i_n ~\rightarrow~& H^i_n ~+~ k
v^i \delta_{n,0}}$$
To avoid using such twisted representations one
usually restricts $v$ to be a weight of $G$, and hence a weight of
$H$.

If a coset state is constructed from highest weight states of $G_k$,
$H_{\ell}$ and $H_{k + \ell}$, then the spectral flow of this state
will not, in general, be represented by highest weight states of the
$G$ and $H$ current algebras (but this state will, of course, be a
highest weight state of the coset model).
 The chiral primary fields given in \cplabels\ are all obtained from
 primary fields of $G$ and $H$.  It turns out (modulo fixed point
problems \LVW\ref\ANS{ A.N. \ Schellekens, ``Field Identification Fixed
Points in $N=2$ Coset Theories'', CERN preprint, TH-6055/91.}) there
are still $Z(G)$ inequivalent spectral flows that map the list
\cplabels\  back into itself, and hence each chiral primary field is
given $Z(G)$ times in \cplabels.  This multiple counting accounts for
the factor of ${1 \over Z(G)}$ in \crdim.

An issue that will be important in the next section is to determine
which elements of the chiral primary ring, $\scr$, can be represented
by $\primh$ with $\lambda_+ = 0$, {\it i.e.}, using only the vacuum in
$H_{g-h}$.  Such fields manifestly form a sub-ring, which we will
denote by $\scf$, of the chiral primary ring.  The elements of $\scf$
are naturally labelled by the highest weights, $\Lambda$, of $G$ at
level $k$.

One finds that if $g-h =1$ in all the simple factors in $H$ then one
can use spectral flow to set $\lambda_+$ to zero.  This is almost
obvious from \cplabels\ and \flow, since a flow by $v = -{ 1 \over
{(g-h)}} [ w(\rho_G) - \rho_G ]$ would accomplish the desired result.
The problem is that the dual Coxeter number, $h$, vanishes in the
$U(1)$ factor of $H$, and so such a $v$ is not necessarily a weight of
$G$, and thus the currents of $G$ would be twisted by such a flow.
However, the only coset models for which one has $g-h=1$ in the simple
factors of $H$ are the $\IC \IP^n$ models, ${{SU(n+1)} \over {SU(n)
\times U(1)}}$, and explicit computations \LVW\ref\Gepb{D. \ Gepner,
\nup{322} (1989) 65.}\Gep\ show that one can always find a
representative, $\Phi^{\Lambda, \lambda_+}_{\lambda_-}$, of each chiral
primary field such that $\lambda_+ = 0$.  Moreover, the set of all such
representatives yields all the chiral primary fields exactly once, {\it
i.e}  none of these representatives is equivalent another under
spectral flow.  As a result, we see that if we only want to consider
untwisted representations of $G$ and $H$, then ${\cal F} \equiv {\cal
R}$ only for the $\IC \IP^n$ models.  Otherwise, ${\cal F}$ is a proper
sub-ring of ${\cal R}$.

\newsec{Topological Matter Models}

To obtain the topological matter model from the $N=2$ supersymmetric
models discussed above, one first ``twists'' the energy momentum
tensor:  $T(z) \rightarrow T(z) + \half \partial J(z)$,
and then uses $G^+(z)$
as a further screening current \ref\Wittop{E. \ Witten \cmp{117} (1988)
353; \cmp{118} (1988) 411; J.M.F. \ Labastida, M. \ Pernici and E.
\ Witten, \nup{310} (1988) 611; D. \ Montano and J. \ Sonnenschein,
\nup{310} (1989) 258.} \ref\moretop{E. \ Witten, \nup{340} (1990) 281;
J. \ Distler, \nup{342} (1990) 523; R. \ Dijkgraaf and E.  \ Witten,
\nup{342} (1990) 486; E. \ Verlinde and H. \ Verlinde, \nup{348} (1991)
457; E. \ Witten, \ijmp{6} (1991) 2775.}\EY\DVV.  From \Htoda\ and
\superuone\ one can easily see that, in the paratoda formulation, this
corresponds simply to replacing $T_{b}^\prime(z)$ of \Htoda\ by:
\eqn\toptoda{T_{b}^{\rm top}(z) ~=~ - \half (\partial \phi )^2 ~+~ i
{}~(\nu_+ {}~-~ \nu_-) \rho_G \cdot \partial ^2 \phi \ ,}
where
$$
\nu_\pm ~=~ \beta_\pm^\prime (k, \ell = g-h) ~=~ \beta_\pm(k, \ell = 0)
{}~=~ \beta_\pm(k,\ell = 0) ~=~ {1 \over \sqrt{k}} ~ \Bigg[ \sqrt{{{(k +
g)} \over {g}}}~ \Bigg]^{\pm 1}\ .$$
Thus the twisted paratoda theory
has precisely the energy-momentum tensor of a $\gkl$ theory but with
$\ell = 0$.  This generalizes the observation in \ref\WLe{W. \ Lerche
\plt{252} (1990) 349.} that when the pure Toda theory is twisted, one
obtains a topological model corresponding to $G_{1,0}$.

We now establish the complete equivalence of the twisted $N=2$
supersymmetric paratoda theory and the topological $G_{k,0}$ coset
models.  Recall that the set of screening currents of the $N=2$ theory
is simply a subset of those that we used for the $\gkl$ theory.
Indeed, the simple roots of $G$ contain one simple root, $\gamma =
\alpha_r$, more than the simple roots of $H$.  The corresponding
``screening'' operators $S_\gamma = \Phi_0^{0, -\gamma}$ and
$S_{-\gamma} = \Phi_{ -\gamma}^{0,0}$ are, in fact, primary fields of
the $N=2$ supersymmetric paratoda theory, and have conformal dimensions
and $Q$-charges:  $h={3 \over 2}$ , $Q= +1$ and $h = \half + {k \over
{2(k+g)}}$ , $Q= -1  + {k \over {(k+g)}}$ , respectively.  This
suggests that we should identify:  $S_{+\gamma}(z) \equiv G^+(z)$ and
$S_{-\gamma}(z) \equiv (\Gminus \phi)(z)$ where $G^\pm(z)$ are the
$N=2$ supercurrents, and $\phi$ is a chiral primary field.  To see that
this identification is correct, consider what the operators $\Phi^{0,
-\gamma}_0(z)$ and $\Phi_{-\gamma}^{0,0}(z)$ represent in the original
$N=2$ coset model.  Let $\lambda^{\pm \bar \alpha}(z)$ be the fermions
of $SO({\rm dim}(G/H)) \equiv H_{g-h}$ and let $J^{\pm \bar \alpha}(z)$
be the currents of $G$ that are not currents of $H$.  The labels, $\bar
\alpha$, are the roots of $G$ that are not roots of $H$.  (See \KS\ for
details.)  The $H_{g-h}$ ground state representation corresponding to
$\lambda_+ = - \gamma$ consists of all the fermions $\lambda^{- \bar
\alpha}(z)$ (and includes $\lambda^{- \gamma}(z)$ itself).  To ensure
that $\lambda_- = 0$, {\it i.e.}, one is in the $H_{k+g - h}$ vacuum,
one must pair these fermions with $J^{+ \bar \alpha}(z)$.  This is
precisely the construction of the supercurrent in \KS.  Similarly, the
primary field with $\Lambda = \lambda_+ = 0$ and $\lambda_- = -\gamma$,
can be represented by the current $J^{- \bar \alpha}(z)$.  Since
$G^\pm(z) ~=~ \sum_{\bar \alpha} \ J^{\pm \bar \alpha}(z) \
\lambda^{\mp \bar \alpha}(z)$, one sees that $J^{- \bar \alpha}(z)$ is
precisely $(\Gminus \lambda^{- \bar \alpha})(z)$.  The fermion
$\lambda^{- \bar \alpha}(z)$ is a representative of the chiral primary
field:
\eqn\cppert{\phi ~=~ \Phi^{0, -\gamma}_{-\gamma}}
with $h =
{ k \over {2(k+g)}}$ and $Q = { k \over {(k+g)}}$.

Thus, if we wish to pass from the $N=2$ theory to the $G_{k, \ell=0}$
theory, we must use the two operators $S_{\pm\gamma}$ as screening
charges.  Using $G^+(z)$ as a screening charge is, of course, already
required in going to the topological version of the $N=2$ theory.  One
can also easily understand the use of $S_{- \gamma}(z)$ as a screening
charge: it is equivalent to perturbing the topological model by
inserting
\eqn\toppert{exp \Big[ - \lambda ~ \int ~d^2z ~G_{-\half}^-
{}~\widetilde G_{-\half}^- ~\phi ~\Big]}
into the topological
correlation functions \foot{The operators $\widetilde G^\pm(\bar z)$
are the anti-holomorphic $N=2$ supercharges.}.
This follows from the
observations of \ref \BNZ{J.\ Bagger, D. \ Nemeschansky and J.-B
\ Zuber, \plt{216} (1989) 320. } \ref\Math{S.D. \ Mathur, ``Quantum
Kac-Moody Symmetry in Integrable Field Theories'', Harvard preprint
HUTMP-90-B-229 (1990).}, in which one can split the integrals of the
form $\int ~d^2z$, into holomorphic and anti-holomorphic contour
integrals, factorize the result and reinterpret the exponential of this
integral as precisely the appropriate generalization of the contour
prescriptions of \ref\Felder{G.~Felder, \nup{317}(1989) 215. }.

The conclusion of all this is that all the correlation functions of
${{G_k \times G_0} \over G_k}$ are equal to the topological
correlations of the $N=2$ supersymmetric HSS models, perturbed by the
top component of the chiral, primarily superfield whose bottom
component is the chiral, primary field $\phi$ of \cppert.

For untwisted representations of affine $G$, the only unitary
representation of $G_0$ is one dimensional and consists of only the
vacuum \ref\Kac{V.G. Kac, {\it Infinite dimensional Lie algebras}, in
Prog. in Math. Vol. 44, Birkh\"auser, Basel 1983.}.  Hence the ${{G_k
\times G_0} \over G_k}$ theory is precisely the $G_k/G_k$ theory
considered in \MSG \witt .  The three point functions in this theory
are precisely the fusion rules of $G_k$.   That is, if $\phi_\Lambda$
is a primary field of $G_k$, then
\eqn\fusion{\phi_\Lambda ~
\phi_{\Lambda^\prime}  ~=~ {N_{\Lambda \,
\Lambda^\prime}}^{\Lambda^{\prime \prime}} ~\phi_{\Lambda^{\prime
\prime}} \ .}
However, we have established this field theory is
isomorphic to the perturbed topological matter model described above,
restricted to the sub-ring $\scf$ of $\scr$.  In particular, the
correlation functions of the two models are equal.

We now recall some of the basic preperties of the $N=2$ supersymmetric
topological matter models \DVV .  The general topological correlation
function is given by:
\eqn\topcor{ \big\langle \phi_{i_1} \ldots
\phi_{i_n} \big\rangle_{\rm top} \equiv ~ \bigg\langle~ \phi_{i_1}
\ldots \phi_{i_n} \ exp\Big[ \sum_\ell t_\ell \int\! d^2z\,
G_{-\half}^- ~\widetilde G_{-\half}^- \phi_\ell (z, \bar z) \Big]
\bigg\rangle \ ,}
where the right-hand-side is computed in the
twisted conformal $N=2$ model.  The two point function, $\eta_{ij} =
\langle \phi_{i}  \phi_{j} \rangle_{\rm top}$ is independent of the
coupling constants, $t_\ell$, and provides a flat metric for the
topological field theory.  The three point function, $C_{ijk}(t)$
$\equiv$ $ \langle \phi_{i}  \phi_{j} \phi_{k} \rangle_{\rm top}$, is
totally symmetric, and satisfies:
\eqn\cprops{\eqalign{{C_{ij}}^m ~
C_{k\ell m} ~=~& {C_{ik}}^m ~ C_{j\ell m} \cr \partial_\ell~C_{ijk}
{}~=~& \partial_i ~C_{\ell jk} \ ,}}
where $\partial_\ell =
\partial_{t_\ell}$, and the indices are raised by $\eta^{ij} =
\eta_{ij}$.  The $n$-point functions can be factorized into products of
the ${C_{ij}}^k$.  For ${t_\ell} = 0$ the three point functions are
nothing other than the multiplication table for the chiral primary
ring, $\scr$.  This ring can be generated by a set of fields, which we
will denote by $x_a$, $a=1, \ldots,M$, and the ring, $\scr$, is
characterized by polynomial vanishing relations $v_b^{(0)}(x_a) = 0$,
$b=1, \ldots,M$ in this ring.  For ${t_\ell} \not= 0$ the fields
$\phi_i(x_a; t_\ell)$ are polynomials in $x_a$ that are
quasihomogeneous in $x_a$ and $t_\ell$.  These polynomials satisfy the
multiplication rule:
\eqn\crmult{\phi_i(x_a; t) ~\phi_j(x_a; t) ~=~
{C_{ij}}^k (t) ~ \phi_k(x_a;t) \quad {\rm mod} \quad \{ v_b(x_a; t) \}
\ ,}
where the $v_b(x_a; t)$ are polynomials in $x_a$, and are the
perturbative vanishing relations.  For a generic perturbation, there
are $\mu$ isolated points $x_a^{(\alpha)}$, that satisfy
$v_b(x_a^{(\alpha)}; t) = 0$.  At these points, the equations:
$$\phi_i(x_a^{(\alpha)}; t) ~\phi_j(x_a^{(\alpha)}; t) ~=~ {C_{ij}}^k ~
\phi_k(x_a^{(\alpha)}; t) \ $$
are satisfied identically.

If the conformal theory has a \LG\ potential, $W_0(x_a)$, then the
perturbed topological matter model has an effective \LG\ potential
$W(x_a;t)$ such that one has $\phi_i(x_a;t) ~=~ -{{\partial W} \over
{\partial t_i}}$, and the partials, ${{\partial W} \over {\partial
x_a}}$, of $W$ yield an equivalent set of vanishing relations to those
provided by the $v_b$.  The points $x_a^{(\alpha)}$ are thus the
critical points of $W$.

Introduce a matrix $D$ such that $D^2 = \eta$.  Since $\eta$ is
symmetric, we can take $D$ to be symmetric (but not necessarily real).
Let $X_i = D C_i D$, where $C_i$ is the matrix ${C_{ij}}^k$.  From
\cprops, one sees that the matrices $X_i$  commute with each other and
are totally symmetric, and so they can be simultaneously diagonalized
by an orthogonal transformation, $\tilde S$.
 Let ${S_i}^\alpha$ be the elements of the matrix  $D \tilde S$, then
$\eta^{ij} {S_i}^\alpha {S_j}^\beta = \delta^{\alpha \beta}$,
${S_i}^\alpha {S_j}^\alpha = \eta^{ij}$, and
\eqn\cdiag{C_{ijk}~S^{j
\alpha} ~S^{k \beta} ~=~ \phi_i^{(\alpha)} {}~\delta^{\alpha \beta}
\ ,}
where $\phi_i^{(\alpha)}$ are the eigenvalues of ${C_{ij}}^k$.
{}From the complete symmetry of $C_{ijk}$ it follows that
$\phi_i^{(\alpha)} = {S_i}^\alpha/{S_0}^\alpha$ and hence:
$$
C_{ijk} ~=~ \sum_{\alpha} ~{{{S_i}^\alpha ~{S_j}^\alpha ~ {S_k}^\alpha
} \over {{S_0}^\alpha }} \ .$$
We may also write this relation as:
\eqn\evmult{\phi_i^{(\alpha)}~\phi_j^{(\alpha)} ~=~ {C_{ij}}^k~
\phi_k^{(\alpha)}}
(with no sum on $\alpha$).  Thus the eigenvalues,
$\phi_i^{(\alpha)}$, are therefore precisely the values of the chiral
primary fields at the points $x_a^{(\alpha)}$, {\it i.e.}
$\phi_i^{(\alpha)} = \phi_i(x_a^{(\alpha)}; t)$.

In the foregoing analysis the set of $\phi_i$ formed a basis for the
entire chiral primary ring, $\scr$.  However, one can consistently
truncate to any sub-ring, $\scf$, of $\scr$ provided that one also
restricts the perturbations in \topcor\ to those that preserve this
truncation.  This is easily done for the sub-ring, $\scf$, introduced
in the last section since the perturbations will preserve the
truncation if and only if the perturbing operator, $G_{-\half}^-
{}~\widetilde G_{-\half}^- \phi (z, \bar z)$, has a representation in the
paratoda theory that consists of some operator in the $G_k$ theory,
tensored with the identity in the $H_{g-h}$ theory.  In particular,
recall that the holomorphic operator $(\Gminus \phi)(z)$ discussed in
the previous section manifestly satisfies this criterion since it was
represented by $\Phi^{0,0}_{-\gamma} (z)$, which has $\lambda_- =0$.
This operator can also be represented by the $G_k$ current
$J^{-\gamma}(z)$.

Thus we can consider the topological matter theory restricted to the
sub-ring, $\scf$, along with the single perturbation \toppert.  For the
right value of $\lambda$ , we have ${C_{\Lambda
\Lambda^\prime}}^{\Lambda^{\prime \prime}}= {N_{\Lambda
\Lambda^\prime}}^{\Lambda^{\prime \prime}}$, where the ${N_{\Lambda
\Lambda^\prime}}^{\Lambda^{\prime \prime}}$ are the fusion
coefficients.  Moreover, from the results of \EV\ we know that the
matrix ${S_\Lambda}^{\Lambda^\prime}$ employed above is precisely the
modular inversion matrix for affine $G$ at level $k$ \Kac:
$${S_\Lambda}^{\Lambda^\prime} ~=~ (i)^{\vert \Delta^+ \vert}
\bigg\vert {{\Gamma^*} \over {(k+g) \Gamma }} \bigg \vert^{-\half}~
\sum_{w \in W(G)}~ \epsilon(w)~ e^{-{{2 \pi i} \over {(k+g)}}
(\Lambda^\prime + \rho_G)\cdot w (\Lambda + \rho_G)}\ ,$$
where
$\vert \Delta^+ \vert$ is the number of positive roots of $G$, $\Gamma$
is the co-root lattice of $G$ and $\Gamma^*$ is its dual.  The
eigenvalues $\phi_\Lambda^{(\Lambda^\prime)}$ of ${C_{\Lambda
\Lambda^\prime}}^{\Lambda^{\prime \prime}}$ are therefore given by:
\eqn\evchar{\phi_\Lambda^{(\Lambda^\prime)} ~=~
{{{S_\Lambda}^{\Lambda^\prime}} \over {{S_0}^{\Lambda^\prime}}}
{}~\equiv~ \chi_\Lambda \Big(e^{-{{2 \pi i} \over {(k+g)}} \Lambda^\prime
\cdot p} \Big) \ ,}
where $p$ is the vector of CSA generators of $G$,
and $\chi_\Lambda$ is the character of the finite $G$ representation
with highest weight $\Lambda$.  The right-hand-side of
\evchar\ constitute the discrete characters of $G$, and have long been
known to satisfy the fusion rules of \EV.

Thus we have established that the values of the $\phi_\Lambda(x_a;
\lambda)$ at the solutions of $v_{\Lambda^{\prime \prime}} (x_a;
\lambda ) = 0$ are given by these discrete characters.  In particular,
one can easily express the $x_a^{(\alpha)}$ in terms of these
characters.

\newsec{Integrable models}

Consider once again the model $M_{k,\ell}(G;H)$. For simplicity we will
specialize to the $N=2$ superconformal  theory, with $\ell=g-h$, but
most of the conclusions of this  section will hold for more general
$\ell$ and $H$. Define operators
\eqn\defspsi{\eqalign{S_{-\psi}(z)
{}~=~& \Phi^{0, \psi}_0 (z) ~\equiv~ {\cal A}^0_{-\psi}(z) ~ exp\big[ - i
\beta_+ ~\psi \cdot \phi \big] \cr S_{\psi}(z) {}~=~& \Phi^{0, 0}_{
\psi} (z) ~\equiv~ {\cal A}^0_{\psi}(z) ~ exp\big[+ i \beta_- ~\psi
\cdot \phi \big] \ ,}}
where $\psi$  is the highest root of $G$.  In
section 2 we indentified $S_{+\gamma}$ and $S_{-\gamma}$ with $G^+(z)$
and $(\Gminus \phi)(z)$, and by virtually identical arguments, we can
identify $S_{-\psi}$ and $S_{+\psi}$
with the conjugate fields $G^-(z)$ and $(\Gplus
\bar \phi)(z)$.  From the properties of hermitean symmetric spaces,
there is an automorphism of $G$ that interchanges $\gamma$ and
$-\psi$.  Indeed if $\alpha_1, \ldots , \alpha_{r-1} , \gamma$ is a
system of simple roots of $G$, then so is $\alpha_1, \ldots,
\alpha_{r-1}, -\psi$ and hence the two systems are  Weyl rotations of
each other. One such automorphism corresponds to the $N=2$ $U(1)$
charge conjugation operation of fields in the coset model.  More
generally there are automorphisms
that mix the roots $\alpha_1,\ldots,
\alpha_{r-1}$ in non-trivial ways with $\gamma$ and $\psi$.

Consider now the $W$-algebra generators of the $G_{k,\ell}$ model.
These can, of course, be represented in the paratoda language, and
moreover the $W$-generators all commute with the screening charges
obtained from the currents $S_{\pm \alpha_i}, i = 1, \ldots ,r $.
(Remember that we have:
$\alpha_r  =  \gamma$. ) If one considers the
$W$-generators in the $G_{k,\ell=0}$ theories, one can then ``untwist''
them to obtain the $W$-generators of the corresponding $N=2$ coset
model.  To give perhaps a better perspective upon this procedure
consider the $N=2$ supersymmetric paratoda theory: $G/\UU \times \UU$.
One can think of this theory as a $G_k$ current algebra, but with a
rescaled torus whose currents have charges at infinity.  One can
presumably write down the $W$-generators for this
model by  using the techniques of
\ref\BBS{F.~Bais, P.~Bouwknegt, K.~Schoutens and M.~ Surridge,
\nup{304} (1988) 348; \nup{304} (1988) 371.}, provided that one
appropriately rescales the torus and adds the proper sub-leading terms.
The resulting $W$-generators will commute with the charges obtained
from  $S_{\pm \gamma}(z)$.  Because the $W$-generators are their own
conjugates, they are invariant under conjugation, and so they must also
commute with the charges obtained from $S_{\pm \psi}(z)$.  This means
that they commute with $G^{\pm}_{-{\half}}$ and with $\oint
(G^-_{-{\half}} \phi ) (z)dz$ and $\oint G^+_{-{\half}} \bar
\phi(z)dz$.  The former means that these $W$-generators are top
components of a superfield, while the latter means that \ref\Zam{{A.B.
Zamolodchikov, {\it JETP Letters} {\bf 46} (1987) 161; ``Integrable
field theory from conformal field theory'' in {\it Proceedings of the
Taniguchi symposium} (Kyoto 1989), to appear in {\it Adv. Studies in
Pure Math.}; R.A.L. preprint 89-001; {\it Int.\ J.\ Mod.\ Phys.}\ {\bf
A4} (1989) 4235.}.} \ref\EG{T.~Eguchi
 and S.-K.~Yang, \plt{224} (1989) 373; \plt{235} (1990) 282.}
\ref\FMVW{P.\ Fendley, S.\ Mathur, C.\ Vafa and N.P. \ Warner,
\plt{243} (1990) 257.},   to first order in perturbation theory, the
$W$-generators provide integrals of motion for the $N=2$ superconformal
model perturbed by
\eqn\intpert{ \lambda \int d^2z
{}~G^-_{-{\half}}\widetilde G^-_{-{\half}}  \phi {}~+~ \bar \lambda \int
d^2 z ~G^+_{-{\half}} \widetilde G^+_{-{\half}} \bar \phi \ \ \  \ . }

The foregoing argument is somewhat heurestic  since it some makes
(highly plausible) assumptions about properties of the $W$-algebra of
the theory.  It also only verifies integrability to first order in
perturbation theory.  For the $\IC \IP^n$ models one can make a much
more complete analysis of the $W$-generators because a simple
superfield formulation is available.  We will see in the next section
that the foregoing argument is born out.

A method for establishing the integrability beyond first order
perturbation theory might be provided by generalizing the techniques of
\EY\EG.  That is,
one can attempt to construct the integrals of motion as
commutators of suitable  multiple integrals of the currents
$S_{-\gamma}$ and $S_{+\gamma}$, the result  should then commute with
the perturbed Hamiltonian.  These techniques appear to work well for
$G=SU(2)$ \EY .

It is elementary to calculate the perturbative corrections to the
superalgebra and establish a Bogomolny bound for the mass of a soliton
in the perturbed theory
\ref\WLNW{W.~Lerche and N.P.~Warner, \nup{358}
(1991) 571.}. One finds,
\eqn\bogobound{ m
\ \ge\ |~\lambda(\omega-1)~ \Delta \phi~|   \ \ , }
where $\lambda$
is the coupling constant, $\omega \ = \ { k \over  k+g}$ is the $N=2$
$U(1)$ charge of $\phi$, and $\Delta \phi \ = \ (  \phi(\sigma =
+\infty) - \phi(\sigma = -\infty ))$ is the difference of the spatial
asymptotic values of $\phi$.  This bound is saturated by chiral
solitons.  If the theory has a \LG formulation then $W(x_a;\lambda)$ is
quasihomogeneous, where $x_a$ has weight $Q_a$, its $N=2$, $U(1)$
charge, and $\lambda$ has weight $1-\omega$.  Hence
\eqn\susyp{W(x_a;\lambda) \ = \ \bigg(~\sum_a Q_a x_a {\del W \over
\del x_a} {}~\bigg) ~+~ (1-\omega)~\lambda ~{\del W \over \del \lambda}
\ \ . }
Since $\lambda$ is the flat coordinate of the perturbation,
we have $ { \del W \over \del \lambda}  =  - \phi $.   Moreover, at
critical points, one has $ {\del W \over \del x_a}  =  0$, and hence
\eqn\delw{ \Delta W \ = \  -(1 -\omega)~\lambda~( \Delta \phi )
\ \ .}
Thus we recover the semi-classical result \ref \OW{D.~Olive and
E.~Witten, \plt{78} (1978) 97.} \FMVW.
\eqn\semicl{m \ \ge \ | \Delta W | \ \ . }

For the $\IC\IP^n$ models, the perturbing field, $\phi$, is equal to
some $\phi_\Lambda$ in $\cal F$.  We can thus express $\Delta \phi$ and
the Bogomolny bound in terms of the character $\chi_\Lambda$.  One
should also note that the effective \LG\ potentials, $W(x_a;\lambda)$,
for the integrable models are not necesseraly first order in
$\lambda$.  Indeed, for $G_k = SU_k(2)$, the potential is a Chebyshev
polynomial\DVV\MSS.
The general form of this  effective potential  can be
obtained by checking that it also reproduces the fusion algebra, and
has also been computed in \Gep\KI.

\newsec{Examples}

The  $N=2$ supersymmetric Grassmanian models:
$${\cal G}_{m,n,k} ~=~
{{SU_k(m+n) \times SO_1 (2mn)} \over {SU_{k+n} (m) \times SU_{k+m}(n)
\times U(1)}}$$
have the property that they are symmetric in $m$,$n$
and $k$.  It is an immediate consequence that they all have three
integrable perturbations (provided that $m$,$n$ and $k$ are distinct).
For $m=1$ and $n=1$, one obtains the type $A$ minimal model and finds
that the integrable perturbation, $\phi$, is the least relevant chiral
primary field.  For $k=1$ and $m=1$ one also obtains the type $A$
minimal models, but this time the integrable perturbation, $\phi$, is
the most relevant, chiral primary field \ref\FLMW{P.\ Fendley,
W.\ Lerche, S.D.\ Mathur and N.P.\ Warner, \nup{348} (1991) 66.}. (It
should be noted that this is not an exhaustive list of the
perturbations that lead to integrable models since it is also known
that perturbing the type $A$ models by the next-to-most relevant field
also leads to an integrable theory \FMVW.)

     The first non-trivial example of a Grassmanian for which the
     analysis of the preceding section implies that there are three
distinct integrable perturbations is provided by the model with $m=3$,
$n=2$ and $k=1$ (or some permutation thereof).  This has a
\LG\ potential:
$$W_0(x_1, x_2) ~=~ \coeff{1}{3} x_2^3 ~+~ 2 x_1^2
x_2^2 ~-~ x_1^4 x_2 \ .$$
Let $W(x_1, x_2; \lambda )$ be the
effective potential after the appropriate
 perturbation has been added.  One then finds the following expressions
for $W(x_1, x_2 ; \lambda) - W_0(x_1, x_2)$:
$$8 \lambda x_1 \, ;
\qquad -4 \lambda (x_2 ~-~ x_1^2)\, ; \qquad 8 \lambda (x_1 x_2 ~+~ 1)
\ ,$$
corresponding to writing the coset model in terms of $G_k =
SU_1(5)$, $SU_2(4)$ or $SU_3(3)$.  (Note that the perturbations have a
charge equal to the level of $G$ divided by $k+g =6$.)

The fusion algebras are obtained by taking $\lambda = 1$.  The
perturbation $-4 (x_2 - x_1^2)$ provides exactly the fusion rules of
$SU_2(4)$, provided that one identifies $x_1$ with the ${\bf 4}$ of
$SU(4)$ and $\half (x_2 + x_1^2)$ with the ${\bf 6}$.  The perturbation
$ 8 (x_1 x_2 + 1)$ of the potential $W_0$  yields the fusion rules of
$SU_3(3)$ provided that one identifies $x_1$ with the ${\bf 3}$ and
$\half (x_2 + x_1^2)$ with the ${\bf 6}$.  Finally, the fusion ring of
$SU_1(5)$ is a sub-ring of dimension $5$, embedded in the chiral ring
$\scr$ whose dimension is $10$.  The ${\bf 5}$ of $SU(5)$ is identified
with $y \equiv \half (x_1^2 - x_2)$, while all the other antisymmetric
tensors of $SU(5)$ map onto powers of $y$.  One also finds that $y^5 =
1$, as one expects.

Another example is provided by the type $D$
modular invariant, minimal
models.  These can be described by taking $G = SO_1(2n+2)$ and $H
=SO_3(2n) \times SO(2)$.  The dimension of $\scr$ is $2(n+1)$.  The
perturbed (integrable) \LG\ potential is:
$$W(x_1,
x_2)~=~\coeff{1}{2n+1} x_1^{2n+1}~-~x_1 x_2^2 ~-~ \lambda x_1 \ .$$
The fusion ring, $\scf$, of $SO_1(2n+2)$ is only four dimensional, and
is isomorphic to $\ZZ_2 \times \ZZ_2$ for $n$ odd and $\ZZ_4$ for $n$
even.  To obtain this ring, one takes $\lambda = 1$ and identifies the
singlet of $SO(2n+2)$ with $1$ and the vector of $SO(2n+2)$ with
$(x_1^{2n} + x_2^2)$.  The two spinors of $SO(2n+2)$ can be identified
with $(x_1^n \pm \alpha x_2)$, where $\alpha = 1$ for $n$ odd and
$\alpha = i$ for $n$ even.

\newsec{$\IC\IP^{n-1}$-models and superfields}

In this section we study the superfield formulation of supersymmetric
$\IC\IP^{n-1}$ models with $G=SU(n)$ and $H=SU(n-1)$.  We  find it
convenient to use $N=1$ superfields rather than $N=2$ superfields.  The
free  superfield formulation of the $\IC\IP^{n-1}$ models can be
obtained from the Lie superalgebra $A(n,n-1)$ through a Hamiltonian
reduction.  Before we study the $\IC\IP^{n-1}$ models we  first review
some basic properties of the super Lie algebra $A(n,n-1)$ that are
relevant for our analysis \ref\supKac{V.G. \ Kac,
{\it Adv. Math. } {\bf 26} (1977) 8.}.

The super Lie algebra $A(n,n-1)$ has a $\ZZ_2$-grading, and therefore
the roots are either even or odd.  If we denote the simple roots by
$\alpha_1,\alpha_2,\ldots,\alpha_{2n-1}, \alpha_{2n} $, then the even
roots are:
\eqn\evenroots{\alpha_i +\alpha_{i+1} + \ldots +
\alpha_{i+2k-2}+ \alpha_{i+2k-1} \ \ \ \ \  k=1,2.\ldots ,\Bigl
[{2n+1-i \over 2 } \Bigr ] \  }
and the odd roots are
\eqn\oddroots{\alpha_i + \alpha_{i+1} + \ldots + \alpha_{i+2k-1} +
\alpha_{i+2k} \ \ \ \ \ \ k =0,1,\ldots , \Bigl [{2n-i \over 2 } \Bigr
] \ \ . }
The simple roots of $A(n,n-1)$ satisfy the following
relations
\eqn\rootrel{\eqalign{ \alpha_{2i-1} \cdot\alpha_{2i} \ & =
\ 1 \cr \alpha_{2i+1} \cdot \alpha_{2i} \ & = -1  \ \ . \cr }}
with
all other inner products being zero (including $\alpha_i \cdot
\alpha_i$).  Next we introduce the fundamental weights $\lambda_1,
\ldots , \lambda_{2n}$ defined by
\eqn\funweig{ \alpha_i \cdot
\lambda_j \ = \ \delta_{ij}   \ \ . }
It is easy to see from \rootrel,
that in terms of the simple roots, the fundamental weights are given
by
\eqn\weightroot{\eqalign{ \lambda_{2i} \ & = \ \alpha_1 + \alpha_3
+ \ldots + \alpha_{2i-3} + \alpha_{2i-1} \cr \lambda_{2i-1} \ & =
\ \alpha_{2i} + \alpha_{2i+2} + \ldots \alpha_{2n-2}+ \alpha_{2n} \ \ .
\cr }}
The super Lie algebra $A(n,n-1)$ contains the even subalgebras
$A_n$ and $A_{n-1}$.  The simple roots of the $A_n$  subalgebra have
the form
\eqn\ansimple{ \alpha_{2i-1} + \alpha_{2i} \ \
\ \ \ \ i=1,\ldots ,n  }
and for $A_{n-1}$ the simple roots re given
by
\eqn\anosimple{ \alpha_{2i} + \alpha_{2i+1} \  \ \ \ \ \  i=1
\ldots,n-1 \ \ .  }
{}From \rootrel\ we see that  the root system for $A_n$ has a positive
definite metric, whereas for $A_{n-1}$, the metric is negative
definite.

After these basic definitions of the Lie super-algebra we are ready to
write down  a free superfield description of the $\IC\IP^{n-1}$
models.  We use  an $N=1$ superfield formulation and therefore
introduce a  single anti-commuting coordinate $\theta$, with $\theta^2
\ = \ 0$.  The super-derivative, $D$, is the square root of the
ordinary derivative:
\eqn\superder{ D \ = \ {\partial \over \partial
\theta } + \theta {\partial \over {\partial \theta}} \ ; \ \ \ \  D^2
\ = \ {\partial \over \partial z } \ \ .}
For the $\IC\IP^{n-1}$ model
we need $2n$ (real) superfields
\eqn\suprfield{ \Phi^i(z,\theta) \ = \ \phi^i(z) + \theta \psi^i(z)
\ \ , }
where $\phi^i(z)$ is a free bosonic field and $\psi^i(z)$ is
a free, real fermion.  The superfields satisfy the following operator
product expansion
\eqn\susyope{\Phi^i(z_1,\theta_1)
\Phi^j(z_2,\theta_2) \ = \ - \delta^{ij} logz_{12} \ \ ,}
where
\eqn\defzot{z_{12} \ = \ z_1-z_2 -\theta_1 \theta_2 \ \ \ . }

The stress tensor is obtained from the the Lax operator \HN \KIto ,
\eqn\laxope{L \ = \ \prod_{i=1}^{2n+1} [ \alpha_0 D -(-1)^i(\lambda_i-
\lambda_{i-1})\cdot D\Phi] \ \ , }
where
\eqn\lamzero{\lambda_0 \ \equiv
\ \lambda_{2n+1} \ \equiv \ 0  \ \ . }
The parameter $\alpha_0$ is
background charge of Feigin-Fuchs representation.  In order to
reproduce the $CP_{n-1}$ models, whose central charge is:
\eqn\cpc{
c \ = \ {3k(n-1) \over k+n} \ \ ,  }
we must  set
\eqn\alphazero{\alpha_0 \ = \ {i \over {\sqrt{ k+n}}}  \ \ . }

In the $N=1$ superfield formulation the stress tensor $T(z)$ is the top
component of an $N=1$ superfield $T(z,\theta)$  with conformal
dimension $3/2$,
\eqn\susyt{ T(z,\theta) \ = \ \half \big( ~G^+(z)
+G^-(z)~\big) + \theta ~T(z) \ \ .}
The fields $G^\pm(z)$ in
\susyt\ are the two supersymmetry generators of the $N=2$ supersymmetry
algebra.  The $U(1)$ current, $J(z)$, of the $N=2$ algebra is the
lowest component of the superfield $J(z,\theta)$
\eqn\jone{J(z,\theta) \ = \ J(z) + \theta~ \half \big(~G^+(z) -
G^-(z)~\big) \ \ . }
{}From the Lax operator we
obtain \HN \KIto\ the currents of the $N=2$ superalgebra for $CP^{n-1}$
model:
\eqn\deftz{\eqalign{ T(z,\theta) \ = &\ -\half
\sum_{i=1}^{2n} \lambda_{2i} \cdot D\Phi^i \alpha_{2i} \cdot \partial
\Phi^i ~-~ \half \sum_{i=1}^{2n} \alpha_{2i} \cdot D\Phi^i \lambda_{2i}
\cdot \partial \Phi^i \cr & \ -{i \over 2 \sqrt{k+n}} \sum_{i=1}^{2n}
\lambda_i \cdot D^3 \Phi \ . }}
and
\eqn\defuones{J(z,\theta) \ =
\ \sum_{i=1}^n \big( \lambda_{2i}\cdot D\Phi \big) \big(\alpha_{2i}
\cdot D\Phi\big) ~-~{i \over \sqrt{k+n}} \sum_{i=1}^n (\lambda_{2i}-
\lambda_{2i-1} ) \cdot \partial \Phi \ \ . }

To fully define the conformal model we need the screening operators.
These are in one to one correspondence with the roots of the Lie
superalgebra $A(n,n-1)$ and its even subalgebras $A_n$ and $A_{n-1}$.
Furthermore, the screening operators commute
with the Lax operator \laxope\
\ref\KatIto{K. \ Ito, ``$N=2$ Superconformal
$\IC \IP_{n}$ Models'', Kyoto
preprint YITP/K-934 (1991).}  and hence they commute  with all
extended symmetries of the $\IC\IP^{n-1}$ model.  From sections 3 and
4, we know that the screening operators that correspond to the
subalgebra $A_{n-1}$ are the ones that play an important role in
determining the integrable perturbation.  (The other screening
operators merely serve to define the parafermionic subsector.) For
completeness we  will write down all the screening operators.  First,
the screening operators corresponding to the roots of $A_n$ have the
form:
\eqn\anscren{ Q_{\alpha_{2i-1}+\alpha_{2i}} \ = \ \int dz
d\theta \ (\alpha_{2i}-\alpha_{2i-1}) \cdot D\Phi e^{{ i \over
\sqrt{k+n}}(\alpha_{2i-1}+ \alpha_{2i} )\cdot \Phi)} \ \ .}
The
screening operators that  are in one to one correspondence with the
roots of the even subalgebra $A_{n-1}$ have the form
\eqn\screanone{Q_{\alpha_{2i}+\alpha_{2i+1}} \ = \ \int dz d\theta \
(\alpha_{2i}- \alpha_{2i+1} )\cdot D\Phi e^{{-{i\over
\sqrt{k+n}}}(\alpha_{2i}+ \alpha_{2i+1} )\cdot \Phi)} \ \ .}
Finally,
associated with every simple root of the Lie subalgebra $A(n,n-1)$, we
have a screening operator
\eqn\screenann{Q_{\alpha_i} \ = \ \int dz
d\theta e^{{i  \sqrt{k+n}} \alpha_i\cdot \Phi} \ \ \ . }

There is a natural $\ZZ_{n+1}$ symmetry of the roots of $A(n,n-1)$ that
map our system of simple roots into a new system.  Under this symmetry
the superfields $\Phi^i$ transform as follows,
\eqn\zntransone{\eqalign{ \alpha_{2i} \cdot \Phi \ & \rightarrow
\alpha_{2i+2}\cdot \Phi  \ \ \ \ \ \ \ i=1,\ldots , n-1 \cr \alpha_{2n}
\cdot \Phi \ & \rightarrow -\lambda_1 \cdot \Phi \ \ \cr }}
and
\eqn\zntranstwo{\eqalign{ \alpha_{2i-1} \cdot \Phi \ & \rightarrow
\alpha_{2i+1} \cdot \Phi \ \ \ \ \ \ \   i=1,\ldots , n-1 \cr
\alpha_{2n-1} \cdot \Phi \ & \rightarrow -\lambda_{2n} \cdot \Phi
\ \ \cr }}
The screening operators  \screanone\ transform under the
$\ZZ_{n+1}$ according to:
\eqn\scretrazn{\eqalign{Q_{\alpha_{2i}+\alpha_{2i+1}} \ \ & \rightarrow
\ \ Q_{\alpha_{2i+2} + \alpha_{2i+3} }  \ \ \ \ \ \qquad \qquad
i=1,\ldots , n-2 \cr Q_{\alpha_{2n-2} + \alpha_{2n-1}}\ \ &\rightarrow
\ \ Q_{\alpha_{2n} - \lambda_{2n}} \ \equiv \ \int dz d\theta \
(\alpha_{2n}+\lambda_{2n}) \cdot D \Phi \ e^{-{i \over \sqrt
{k+n}}(\alpha_{2n} -\lambda_{2n})\cdot \Phi}   \cr Q_{\alpha_{2n} -
\lambda_{2n}} \ \ & \rightarrow \  \ Q_{\alpha_{1} - \lambda_{1}}
\ \equiv \ - \int dz d\theta \ ( \lambda_1 + \alpha_{1}))\cdot D\Phi
e^{{-{1 \over \sqrt{k+n}}}(\alpha_1
 - \lambda_1 )\cdot \Phi } \cr Q_{\alpha_{1}-\lambda_{1}} \ \ &
 \rightarrow \ \ Q_{\alpha_{2} + \alpha_{3} }  \ . \cr}}
The roots,
$\alpha_{2i-1} + \alpha_{2i}$, $i = 1, \ldots, n-1$, $\alpha_{2n} -
\lambda_{2n}$ and $\lambda_1 - \alpha_1$, which appear in the foregoing
transformations, in fact define a system of simple roots for $\hat
A_{n}$.

In the transformations of the screening operators, we introduced two
new operators:
\eqn\defnewop{\eqalign{ &\int d\theta ~(\alpha_{2n} +
\lambda_{2n}) \cdot D\Phi ~ e^{{-{i \over \sqrt{k+n}}} ( \alpha_{2n}-
\lambda_{2n} )\cdot \Phi } \cr & \int d\theta ~
(\lambda_{1}+\alpha_1)\cdot D\Phi ~ e^{{-{1 \over
\sqrt{k+n}}}(\lambda_1  + \alpha_1 )\cdot \Phi } \ . \cr}}
{}From the operator product expansion of these fields with stress tensor
we find
that the conformal dimension of both of these fields is $(2k+n)
/2(k+n)$.  The operator product of these operators with the $U(1)$
current of the $N=2$ algebra shows that these fields have opposite
$U(1)$ charge.  These fields correspond to the operators $S_{-\gamma}$
and $S_{\psi}$ that were discussed in section 4.

One can use the explicit realization of the super $W$-generators
provided by \laxope\ to verify that the top component of the $N=2$
supermultiplet are invariant, up to total derivatives, under the
$\ZZ_{n+1}$ symmetry.  As a result the charges associated with the
generators commute with the perturbation.
Hence, to first order in this perturbation, the top components
of the $N=2$ super $W$-algebra can be
extended to integrals of motion for the perturbed model.
(This argument is a straightforward generalization of the one
that was used in \EG\ and \FLMW\ to estabish similar results
for purely bosonic formulations of $W$ and super-$W$ algebras.)

It is relatively easy to identify the affine super-Toda action that
underlies the integrable model.  Motivated by the purely bosonic
theory, one pairs a holomorphic screening charge (or a relevant
perturbation) with its anti-holomorphic counterpart to obtain the
following  potential term of a super-Toda action:
\eqn\supertoda{\eqalign{& \sum_{i=1}^{n-1} \int d^2 z d \theta d \bar
\theta (\alpha_{2i}-\alpha_{2i+1}) \cdot D \Phi
(\alpha_{2i}-\alpha_{2i+1}) \cdot \bar D \Phi e^{{- i \over \sqrt
{k+n}}\alpha_{2i}+\alpha_{2i+1}\cdot  \Phi} + \cr & \int d^2 z d \theta
d \bar \theta (\alpha_{2n}+\lambda_{2n}) \cdot D\Phi
(\alpha_{2n}+\lambda_{2n}) \cdot \bar D\Phi e^{{-i \over
\sqrt{k+n}}\alpha_{2n}-\lambda_{2n} \cdot \Phi} + \cr & \int d^2
d\theta d \bar \theta (\alpha_{1}+\lambda_{1}) \cdot D\Phi
(\alpha_{1}+\lambda_{1}) \cdot \bar D\Phi e^{{ -i\over \sqrt{k+n}}
\alpha_{1}-\lambda_{1} \cdot \Phi} \ , \cr}}
where $\Phi \ =
\ \Phi(z, \theta, \bar z, \bar \theta)$.  One can verify, using the
techniques of \BNZ \Math, that when employed in a perturbation
expansion, the foregoing Toda potential generates the appropriate
screening for the conformal sector, along with the proper  insertions
of the relevant perturbations.

To conclude this section we examine the action of $\ZZ_{n+1}$ symmetry
of the other screening operators.  For the screening operators
\anscren\ we have the following transformation law
\eqn\antran{\eqalign{ Q_{\alpha_{2i-1}+\alpha_{2i}} & \ \rightarrow \
Q_{\alpha_{2i+1}+\alpha_{2i+2}} \ \ \ \ \ i=1,2, \ldots , n-1 \cr
Q_{\alpha_{2n-1}+\alpha_{2n}} & \rightarrow \int dz d\theta
(\lambda_{2n}-\lambda_1)\cdot D\Phi e^{-{i\over
\sqrt{k+n}}(\lambda_1+\lambda_{2n})\cdot \Phi} \equiv Q_{\lambda_1+
\lambda_{2n}} \cr Q_{\lambda_1+\lambda_{2n}}
 & \rightarrow Q_{\alpha_1 +\alpha_2} \ .  \cr}}
(It is interesting
to note that sum of the weights $\lambda_1+\lambda_{2n}$ appearing in
\antran\ is just the sum of the simple roots of the super Lie algebra
$A(n,n-1)$.)
 From \antran\ we see that $\ZZ_{n+1}$ transformation generates  a new
operator
\eqn\annew{\int d\theta (\lambda_{2n}-\lambda_1) \cdot D \Phi
e^{{i \over \sqrt{k+n} }(\lambda_1+\lambda_{2n})\Phi} \ \ . }
 From the operator product of this operator with
the stress tensor $T(z)$ and the $U(1)$ current  we find that the
conformal dimension of the field \annew\ is $ {k+n+1
 \over k+n} $ and the $U(1)$ charge is zero.  This operator is
irrelevant and therefore does not give rise to an interesting perturbed
conformal field theory.

The $\ZZ_{n+1}$ symmetry acts on the operators \anscren\
according to:
\eqn\finalzn{\eqalign{Q_{\alpha_{2i-1}}\ &   \rightarrow \
Q_{\alpha_{2i+1}} \ \ \  i=1,2, \ldots n-1 \cr Q_{\alpha_{2n-1}}  &
\ \rightarrow \ \int dz d\theta e^{ - i\sqrt{k+n} \lambda_{2n} \cdot
\Phi } \equiv Q_{\lambda_{2n}} \  \cr Q_{\lambda_{2n}} & \ \rightarrow
Q_{\alpha_1} \cr }}
and
\eqn\annnne{\eqalign{Q_{\alpha_{2i}} \ &
\rightarrow \ Q_{\alpha_{2i+2}} \ \ \ i=1,2,\ldots, n-1 \ \cr
Q_{\alpha_{2n}} \ & \rightarrow \ \int dz d\theta e^{-i\sqrt{k+n}
\lambda_1 \cdot\Phi } \equiv Q_{\lambda_1}  \cr Q_{\lambda_1} \ &
\rightarrow  Q_{\alpha_2} \ \ \  . \cr }}
{}From equations \finalzn\ and \annnne\ we
see that the $\ZZ_{n+1}$ symmetry generates two new operators,
\eqn\gennew{\eqalign{ & \int d\theta e^{-i\sqrt{k+n} \lambda_{2n} \cdot
\Phi }\ \ , \cr & \int d \theta e^{-i \sqrt{k+n} \lambda_1 \cdot \Phi
}\ \ . \cr }}
As before  the operator product with the stress tensor shows that
these also correspond to irrelevant operators and therefore do not give
rise to any interesting perturbations.

\newsec{Conclusions}

Using the paratoda description of $N=2$ supersymmetric models we have
established the connection betweeen the coset models $G_k/G_k$ and
twisted $N=2$ superconformal models.  To obtain the equivalence of
these models we saw that we have to perturb the topological $N=2$ model
by the top component of a chiral primary superfield, whose bottom
component is a  very particular chiral primary field (defined in
\cppert) with conformal dimension ${k \over 2(k+g)}$.  Once this
perturbation is made, the fields of the $G_k/G_k$ theory can be
identified as a sub-ring, $\cal F$, of the chiral ring, $\cal R$, of
the perturbed $N=2$ theory.  The correlation functions of the $G/G$
theory are equal to the topological correlators of the fields in the
sub-ring, $\cal F$, of $\cal R$ in the $N=2$ theory, and, in
particular, the fusion coefficients of
$G$ can be identified with the three
point functions of the perturbed $N=2$ theory.

We have also shown that the chiral primary fields,
$\Phi_\Lambda(x_a;\lambda)$, when evaluated at the points
$x_a^{(\alpha)}$ that solve the vanishing relations, are nothing other
than the discrete characters of $G_k$.

As has been conjectured elsewhere \CeVa , we also find that the
perturbation that  leads to the $G_k/G_k$ theory is also precisely the
one that leads to integrable models.  Once again, the paratoda
formulation leads to a simple understanding of the integrals of motion
of these perturbed conformal field theories.

There remain several interesting questions.  We have, for simplicity,
taken the diagonal modular invarinat for $G$.  The result that we have
established here is valid for any choice of modular invariant for
$G_k$.  Thus one can take the attitude that this paper defines
$G_k/G_k$ theories for special modular invariants by relating them to
perturbed, twisted $N=2$, superconformal theories.  It would be
interesting  to understand whether  there is some fusion algebra
interpretation of $G_k/G_k$ theory with a special modular invariant.
In view of the results in the last section of \ref\LeWa{W.~Lerche and
N.P~Warner ``Solitons in Integrable, $N=2$ Supersymmetric Landau
Ginzburg Model'', to appear in the proceedings of the
{\it Strings and Symmetries Conference},
Stony Brook, May 1991. }, the modular
invariants for $SU_k(2)$ will almost certainly provide some interesting
ring structures.

Finally, we have seen that the fusion ring is naturally a sub-ring of
${\cal R}$.  However, all the fields in $\cal R$ can be interpretted in
terms of ${G_k \times G_0 \over G_k}$ provided one twists $G_0$.  It
might be interesting to study such extensions of the $G_k /G_k$
theory.

\bigskip

\noindent{\bf Acknowledgements}
\smallskip

We would both like to thank W. Lerche for extensive
and valuable discussions about
$N=2$ superconformal theories, $N=2$ super-$W$ algebras, and
topological field theories.  D.N would like to thank S. Yankielowicz
for discussions on $G_k/G_k$ models.

\vfill \listrefs \eject \end